# DW&C:Dollops Wise Curtail IPv4/IPv6 Transition Mechanism using NS2.

**[1]Dr.Hanumanthappa .J., [2]Mr.Annaiah .H**

[1]Dept of Studies in CS,Manasagangothri,University of Mysore,Mysore-570006,hanums_j@yahoo.com
[2]Dept of CS&Engg,Govt Engg College,Hassan,Karnataka,India.
annaiahh@gmail.com

***Abstract***—BD-SIIT and DSTM are widely deployed IPv4/IPv6 Transition mechanism to improve the performance of the computer network in terms of Throughput,End to End Delay(EED) and Packet Drop Rate(PDR).In this journal paper we have Implemented and Compared the Performance Issues of our newly proposed Dollops Wise Curtail(DW&C)IPv4/IPv6 Migration Mechanism with BD-SIIT and DSTM in NS2. Implementation and Comparison Performance Analysis between Dollops Wise Curtail,BD-SIIT and DSTM shows that Dollops Wise Curtail IPv4/IPv6 migration algorithm performance outperforms than BD-SIIT and DSTM.Based on extensive simulations,we show that DW&C algorithm reduces the Packet Drop Rate(PDR),End to End Delay(EED) and achieves better Throughput than BD-SIIT and DSTM.In our research work observation,the performance metrics such as Throughput,EED and PLR for DW&C,BD-SIIT and DSTM are measured using TCP,UDP,FTP and CBR Traffics.

## I.Introduction

Divide and Conquer is an important data structure technique which is mainly used to implement wide range of problems such as Transition issues/Routing issues,Mobility issues,Sensor Networking issues,MANET's issues,Performance Analysis of Protocols issues etc[1][24][30].The Divide-and-Conquer paradigm is mainly used to solve efficient and effective algorithms in IPv4/IPv6 migration mechanism.The Divide and Conquer technique plays a important key role in all different types of sorting/searching mechanisms such as Heap Sort,Selection Sort,Bubble Sort,Radix Sort,Quick sort,Merge Sorting techniques and Binary Search,Linear Search,Ternary Search etc.Divide-and-Conquer algorithms are naturally designed and implemented as recursive procedures[24][25].Divide and conquer algorithms can also be implemented by a non-recursive algorithm that stores the partial sub-problems in some explicit data structure,like a stack,queue, priority queue,double ended queue(Deque) to design the Dollops Wise Curtail(DW&C).The Dollops Wise Curtail(DW&C)implemented in this way is also called basic DW&C[30].The Dollops Wise Curtail can also be implemented in software using NS2/OPNET++/OMNET++ Simulator etc[24][25].

## II.Related Works

### 2.1.BD-SIIT

In this research work we have also proposed a new transition algorithm called BD-SIIT.The BD-SIIT(Stateless Internet Protocol/Internet control messaging Protocol Translation(SIIT) is a novel IPv6 transition mechanism which allows IPv6 only hosts to speak to IPv4 only hosts.The BD- SIIT translator is one which is recognized as a stateless IP/ICMP translation,because the translator executes each translation individually without any reference to previously converted packets[1][24][25].

### 2.2.DSTM

The Dual stack transition mechanism(DSTM) is also mainly created to make IPv4 to IPv6 migration for an IPv6 dominant network which consists of both IPv4 and IPv6 configured links and can communicate with IPv4 and IPv6 nodes[12][13].The DSTM is an IPv4 to IPv6 transition proposal based on the uses of IPv4 over IPv6 dynamic tunnels,and the temporary allocation of IPv4 global addresses to dual stack hosts.The DSTM is intended for IPv6 only networks in which hosts need to exchange information with other IPv4 hosts or applications[28].

We can find a couple of works carried out in this direction.Hanumanthappa.J,Manjaiah.D.H.(2006) have implemented Divide and Conquer based DST/List Ranking/Slice Wise Curtail based IPv4/IPv6 Transition Mechanism.They also compared Performance Analysis of Divide and

Conquer based DST/Slice Wise Curtail /List Ranking technique with our previously implemented BD-SIIT and DSTM. The simulation results observed for SL&C,BD-SIIT and DSTM are clarified that SL&C works better than BD-SIIT and DSTM in terms of Throughput,EED and PDR[30].

Hira Sathu,Mohib A Shah(2009) have heuristically tested the three well known IP Transition techniques such as IPv6to4,IPv6in4,and Dual Stack. In their practical experiment two tunneling and a Dual Stack mechanisms were configured and repercussion of these mechanisms on Video Packets was observed. Comparison between video protocols illustrates that MPEG-2 was highly repercussion by tunneling mechanisms having almost the same amount of bandwidth wasted while MP4 was least repercussion by tunneling mechanism.

Eun-Young Park,Jae-Hwoon Lee et al(2009) have proposed and implemented the IPv4/IPv6 Dual Stack Transition Mechanism(4 to 6 DSTM). Which can work even in the case that hosts in the IPv4 network initiate connections with hosts in the IPv6 network[2][8].

Chiranjit Dutta and Ranjeet Singh(2012) have tested and heuristically reckoned two transition mechanisms namely IPv6/ IPv4 Tunneling and Dual-stack mechanism,as they relate to the performance of IPv6.They explore the repercussion of these approaches on end-to-end user application performance using metrics such as Throughput,Latency and Host CPU utilization. All experiments were deportmented using Three dual stack(IPv4/IPv6) routers,an IPv6 router and two ends–stations running Windows 7,loaded with a IPv4/IPv6 dual stack.

The structure of this research paper is as follows. In section 2 the proposed methodology is explained with a neat block diagram along with diminution tree. The Performance analysis methods and metrics are described in section 3.The simulation results are discussed in section 4,the section 5 illustrates discussion and observations and finally the paper is concluded in section 6.

### III.Proposed Methodology

In this paper we have proposed and Implemented Dollops Wise Curtail(DW&C) based IPv4/IPv6 Transition Mechanism using Divide and Conquer strategy[30].One more Divide and Conquer methodology entitled Dollops-Wise Curtail(DW&C) is successfully installed to detect the IPv4/IPv6 transition performance issues in a proper manner. The block diagram of the newly proposed Divide and Conquer based DW&C Transition of IPv4 to IPv6 algorithm is shown in Fig.1.Which clearly depicts the different junctures. The steps are:(a)Divide and Conquer(D&C) based on Dollops Wise Curtail(DW&C) BD-SIIT technique[30].

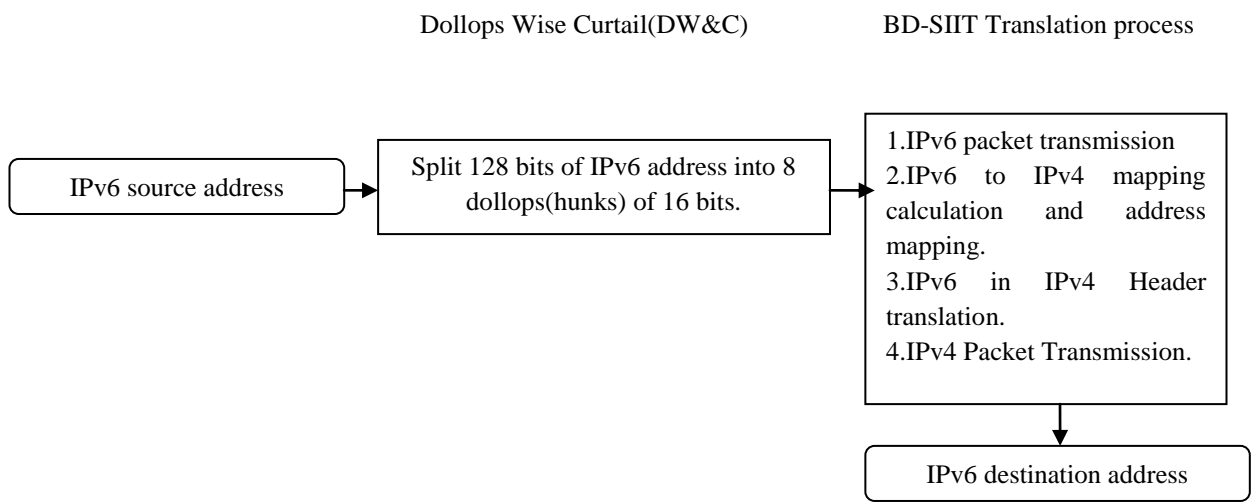

"Fig.1":Block Diagram of the Proposed Methodology(DW&C).

The Dollops Wise Curtail(DW&C) is started by dividing the 128 bits of IPv6 address into an 8 Dollops(pieces) called hunks[30].The size of each hunk is 16 bits which is equivalent to one word size.In the research work,hunks are considered from the Left to

Right[30].The research work fixes the width of each hunk is 16 bits. However,with a clear observation it was mentioned that the size of the Dollops should be always 16 bits. Since the IPv6 address is divided into 8 Dollops from left to right direction[29][30].The experimental research work also clarified that the mechanism of traversing an 128 bits source IPv6 address does not affect on the performance of the Dollops(Slab) Wise Curtail(DW&C) algorithm for IPv4/IPv6 transition mechanism[29][30].

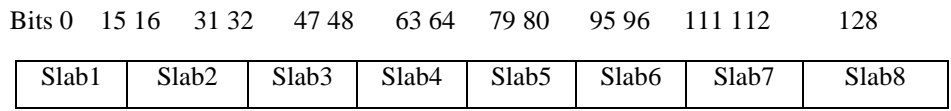

**"Fig.2":Slicing(Slabing) of the 128 bits IPv6 address into 8 stripes(hunks)**

*A.Working Methodology of Dollops Wise Curtail(DW&C)*

The proposed algorithm is called to be Dollops Wise Curtail addresses and it is illustrated in Fig.3. The basic structure of this algorithm is called as succession(position) based tree which is designed by using Piece(Slabs) Classification[29][30].The destination IPv6 address of the incoming packet is divided into 8 separate 16-bit long hunks(stripes) accordingly[29][30].

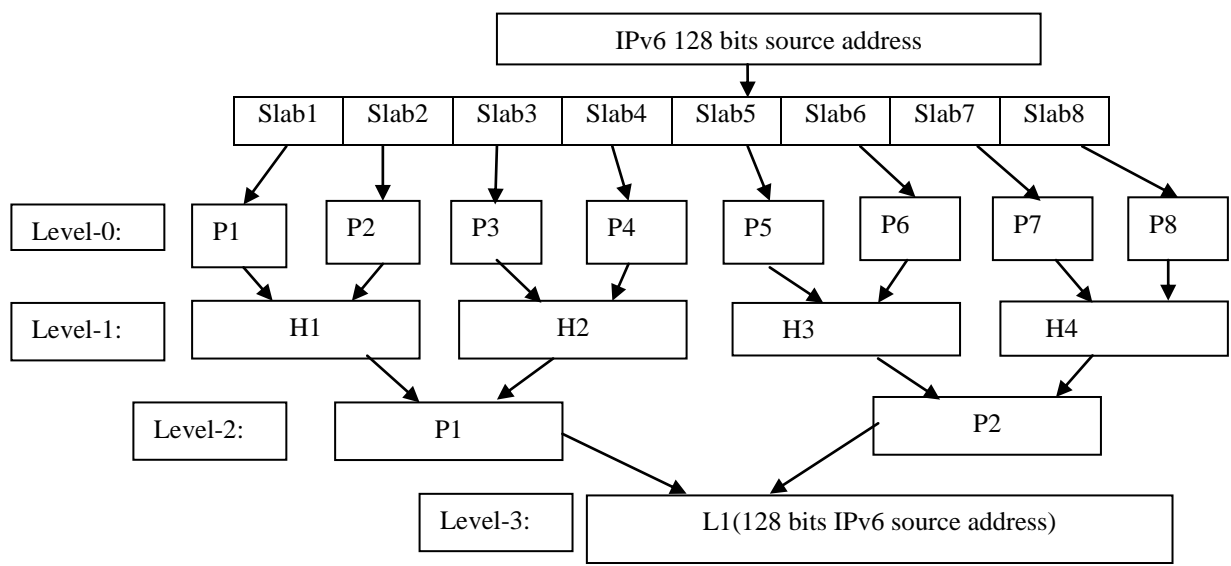

**"Fig.3":Decomposition of an IPv6 128 bits address into 8 Dollops based on 16 bits word length components.**

```
The Pseudo Code Of Dollops Wise
Curtail(DW&C) IPv4/IPv6 Transition
struct dwc{unsigned char ml;
unsigned int strt;
unsigned int stp;
        }dwc[8];
ml=total_size_of_IPv6_address;
div=ml/16;
md=ml%16;
for(z=0;z<div;z++)         {
dwc[z].strt=prx[z].stp=prx[z];
dwc[z].ml=16;}
  dwc[div].strt=prx[div]>>md<<md;
  dwc[div].stp=(dwc[div].start)/(((1<<(md+1))-1);
  dwc[div].ml=md;
 for(z=div+1; z<8; z++)    {
 dwc[z].strt=0;
 dwc[z].stop=0xFFFF;
 dwcl[z].ml=0;    }
prx[z]:the ith 16-bit value of an IPv6 address space.
Pseudocode-1:The Pseudo code of  Dollops
division for 128 bits of IPv6 address.
```

## IV. Performance Analysis Methods And Metrics

Performance analysis is an important criterion for designing both IPv4 and IPv6 network. The metrics we considered in this innovative research work are latency, throughput and Packet Drop Rate(PDR). The Discrete event simulator NS2 particularly popular in the performance analysis of protocols community is mainly considered as the common scaffolding to simulate the various existing and the proposed IPv4/IPv6 transition scenarios in the research work. NS2 simulates the IPv4/IPv6 transition performance queries by using the various performance parameters such as Throughput, End-to-End Delay(RTT)(Latency), Packet Loss(Drop)Rate and it presents one of the best case scenarios for the typical use of IPv4/IPv6 transition. The Implementation of IPv4/IPv6 BD-SIIT translator is simulated by using NS2 simulator.

*A. Performance Metrics*

*In order to evaluate the performance issues of DW&C, BD-SIIT and DSTM IPv4/IPv6 Transition mechanism we deploy three different categories of key parameters such as Throughput, End-to-End Delay, Jitter and Packet Drop Rate[21][26].*

Throughput: Throughput is a ratio of the number of bits received at the destination work stations to the total simulation time[21][26].

Throughput=($\sum$ Packets received x packets size x 8 % total simulation time)x100 % ----------(i)

Average End-to-End Delay: This is a difference between when the Packet ends it time and when the Packet starts it time.

Delay=$\sum$(Packet End Time – Packet Start Time)-(ii)

Packet Loss Rate(PLR)/Packet Drop Rate(PDR): Packet loss occurs during the communication between two or more hosts across the network. When two hosts exchange packets between their operating systems and some of the packets get dropped during the transmission due to overload which is called Packet Loss Rate. Most commonly, the packet gets dropped before the destination can be reached[21][26].

Packet Loss Rate/dropped rate(PLR/PDR)=$K_s - K_r$(iii)

Where $K_s$ is the amount of packet sent at source and $K_r$ is the amount of packet received at destination.

Jitter: Jitter is one which is defined as fluctuation of end to end delay from one packet to a next connection flow packet[21][26].

Jitter (J)= $O_{i+1} - O_i$ -------------(iv)

Where $O_{i+1}$ is Obstruction of i+1th packet and $O_i$ is the Obstruction of ith communication packet.

## V. Simulation Results

We compared the Performance of D&C based DW&C with two different transport layer protocols TCP and CBR. In the experimental scenario we have compared the Performance Issues of D&C based DW&C with BD-SIIT, DSTM IPv4/IPv6 Transition mechanisms[21][26]. The experiment it consists of varying packet size of each datagram i.e 32,64,128,256,512,768,1024 bytes and the Maximum Queue Length(IFQ) is set to 50 packets etc. Each simulation executes for 200 sec of simulation time. The number of runs by varying the traffic and the researcher has computed the Throughput, Average End to End delay and Packet Drop Rate etc[21][26].

**"Table 1": Average End-to-End Delay for IPv6 network with DSTM.**

| Sl.No | Packet Size(Bytes) | FTP End-to-End delay(ms) | CBR End-to-End delay(ms) |
|---|---|---|---|
| 01 | 256 | 90 | 70 |
| 02 | 512 | 100 | 85 |
| 03 | 1000 | 105 | 93 |
| 04 | 1256 | 130 | 122 |

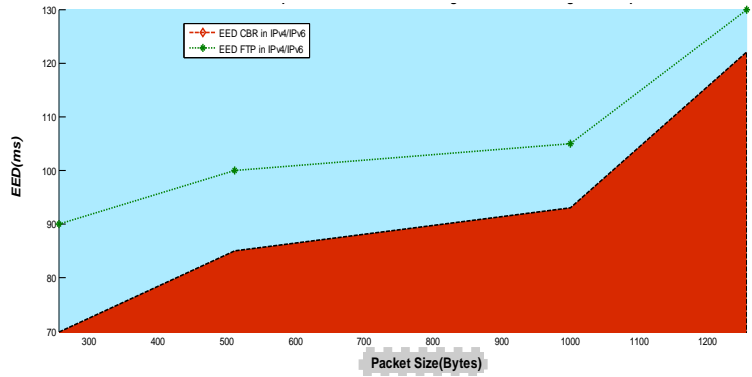

**"Fig.4":Calculation of Average(Mean)End-to-End Delay(EED) using FTP and CBR using an area graph.**

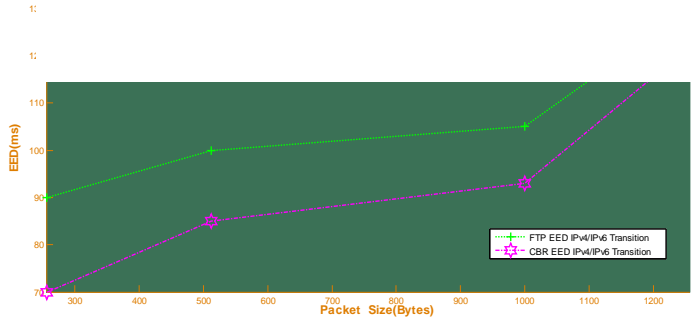

**"Fig.5":Computation of DSTM mean End-to-End Delay(EED) using FTP and CBR with line graph.**

The Table-2 shows the Average EED of DW&C and BD-SIIT using FTP and CBR.The Fig.4. shows the comparison of EED of DW&C,BD-SIIT using FTP and CBR and the Fig.5. shows the comparison of EED of DSTM,DW&C and BD-SIIT using FTP and CBR[11][19][21][26].

**Table 2:Mean EED of DW&C and BD-SIIT using FTP and CBR.**

| Sl.No | Packet Size(Bytes) | EED for DW&C using FTP(ms) | EED for DW&C using CBR(ms) | EED for BD-SIIT using FTP(ms) | EED for BD-SIIT using CBR(ms) |
|---|---|---|---|---|---|
| 01 | 256 | 80 | 60 | 85 | 63 |
| 02 | 512 | 90 | 65 | 94 | 68 |
| 03 | 1000 | 94 | 82 | 98 | 89 |
| 04 | 1256 | 125 | 94 | 127 | 96 |

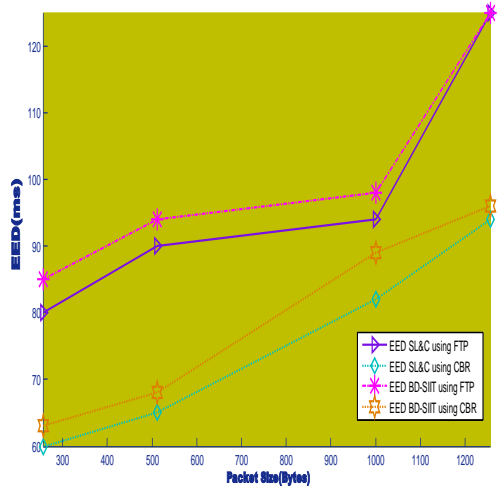

**"Fig.6":Comparison of EED of DW&C,BD-SIIT using FTP and CBR.**

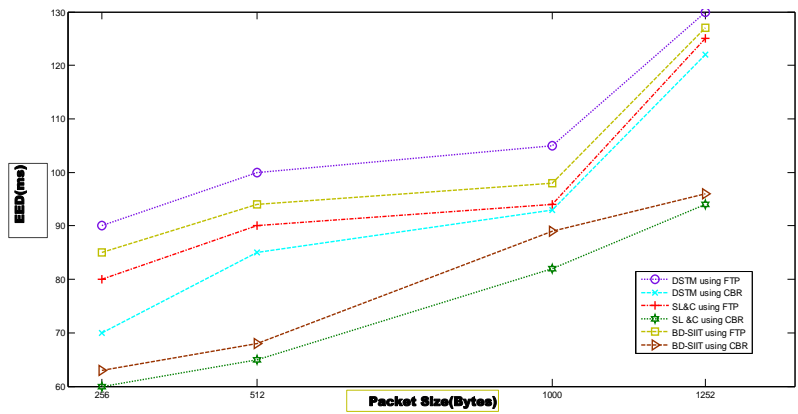

**"Fig.7":EED Comparison of DSTM,DW&C,BD-SIIT using FTP and CBR.**

The Table-3 shows the Throughput comparison for DW&C,BD-SIIT and DSTM.

**"Table 3":Comparison of DW&C,BD-SIIT and DSTM Throughput.**

| Sl.No | Packet Size(Bytes) | Throughput of DW&C(%) | Throughput of BD-SIIT(%) | Throughput of DSTM(%) |
|---|---|---|---|---|
| 01 | 32 | 98.8 | 97.0 | 94.1 |
| 02 | 64 | 96.1 | 95.6 | 92.6 |
| 03 | 128 | 94.7 | 94.0 | 90.9 |
| 04 | 256 | 88.4 | 84.7 | 87.1 |
| 05 | 512 | 86.8 | 82.3 | 85.6 |
| 06 | 768 | 82.3 | 77.8 | 80.2 |
| 07 | 1024 | 76.6 | 74.2 | 74.5 |

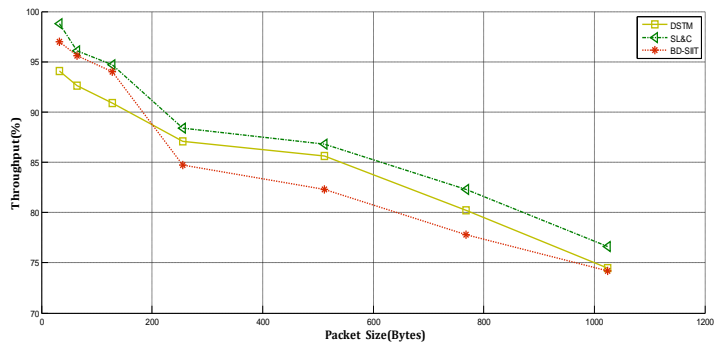

"Fig.8":Comparison of Throughput of DWC,DSTM,BD-SIIT transition mechanism using line graph.

"Table-4":Calculation of PLR of DW&C,BD-SIIT and DSTM when the packet size varies from 32 bytes to 1024 bytes using TCP traffic.

| Sl.No | Packet size(Bytes) | PLR of DW&C(%) | PLR of BD-SIIT(%) | PLR of DSTM(%) |
|---|---|---|---|---|
| 01 | 32 | 0.10 | 0.23 | 0.34 |
| 02 | 64 | 1.15 | 1.28 | 1.39 |
| 03 | 128 | 1.34 | 1.42 | 1.65 |
| 04 | 256 | 1.65 | 1.98 | 2.79 |
| 05 | 512 | 2.36 | 2.94 | 3.41 |
| 06 | 768 | 2.68 | 3.46 | 3.98 |
| 07 | 1024 | 4.58 | 4.80 | 4.94 |

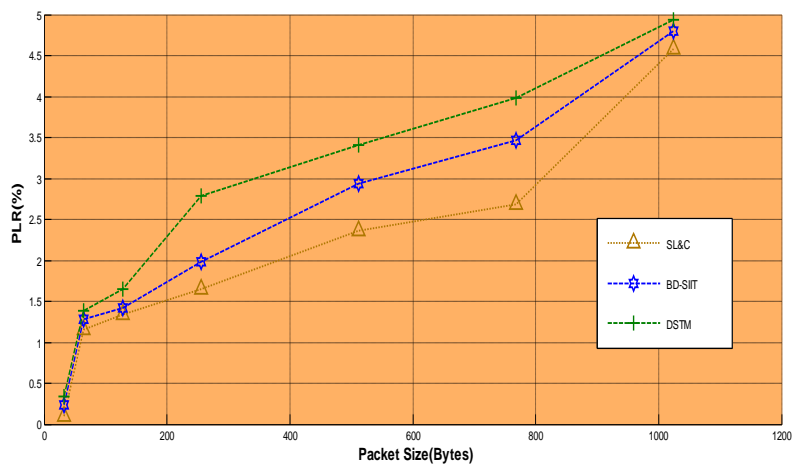

"Fig.9":Comparison of Throughput of DW&C,DSTM,BD-SIIT transition mechanism using line graph.

### VI.Discussion and Observations

This section,now presents and discusses the results of our research work based on Dollops Wise Curtail(DW&C) IPv4/IPv6 transition mechanism.The results observed for DW&C,BD-SIIT and DSTM are showed that DW&C works better than BD-SIIT and DSTM in terms of Throughput,EED and PLR. In the first observation,the performance metrics such as Throughput,Average EED and PLR for DW&C,BD-SIIT and DSTM are measured using TCP and UDP Traffic. In the second observation the performance metrics such as Throughput,Mean EED,and PLR for DW&C,BD-SIIT and DSTM are measured using FTP and UDP protocols[11][19][21][26].

## VII.Conclusions

The purpose of research work,was to identify the performance parameters of DW&C,BD-SIIT and DSTM using TCP,UDP,FTP and CBR protocols. By empirically simulating NS-2 on the test-bed,the following conclusions are obtained as follows:

i.The outcome for Throughput indicates 98.8(%) when the packet size was 32 bytes and it varies up to 76.6(%) when the packet size was 1024 bytes. The Throughput of BD-SIIT varies from 97.0(%) to 74.2(%) and DSTM Throughput varies from 94.1(%) to 74.5(%) when the packet size increases from 32 bytes to 1024 bytes[11][19].

ii.End-to-End delay(EED) experienced in all three transition mechanism show that DW&C delay varies from 80(ms) to 125(ms) for FTP Traffic and 60(ms) to 94(ms) for CBR,the BD-SIIT delay varies from 85(ms) to 127(ms) for FTP and 63(ms) to 93(ms) for BD-SIIT when the packet size varies from 256 to 1256 bytes. The research work was observed least delay in DW&C and maximum delay in BD-SIIT[21][26].

iii. The Packet Loss Rate(PLR) observed on all the three different transition mechanism proves that DW&C PLR varies from 0.10(%) to 4.58(%),BD-SIIT PLR varies from 0.23(%) to 4.80(%) and DSTM PLR varies from 0.34(%) to 4.94(%) for TCP and UDP traffic and it rectify that performance of DW&C is better than BD-SIIT and DSTM in terms of PLR. By using FTP and CBR Traffic the DW&C PLR varies from 1.67(%) to 5.12(%) for FTP,1.85(%) to 5.64(%) for CBR,the BD-SIIT PLR varies from 1.72(%) to 5.22(%) for FTP and 1.98(%) to 5.75(%) for CBR,the DSTM PLR varies from 1.86(%) to 5.42(%) for FTP and 2.00(%) to 5.82(%) for CBR when the packet size varies from 256 to 1256 bytes[19][21][26].